\documentclass[]{spie}  %>>> use for US letter paper
\usepackage{amsmath,amsfonts,amssymb}
\usepackage{graphicx}
\usepackage[colorlinks=true, allcolors=blue]{hyperref}
\usepackage{gensymb}

\title{Integrated modeling of wavefront sensing and control for space telescopes utilizing active and adaptive optics}

\author[a]{Kevin Z. Derby}
\author[a]{Kian Milani}
\author[a]{Solvay Blomquist}
\author[a]{Kyle Van Gorkom}
\author[a]{Sebastiaan Haffert}
\author[a]{Hyukmo Kang}
\author[a]{Hill Tailor}
\author[a,b]{Heejoo Choi}
\author[c]{Christopher B. Mendillo}
\author[d]{Jared R. Males}
\author[a,b,d]{Daewook Kim}
\author[d]{Ewan S. Douglas}

\affil[a]{Wyant College of Optical Sciences, University of Arizona, Tucson, AZ, USA}
\affil[b]{Large Binocular Telescope Observatory, University of Arizona, Tucson, AZ, USA}
\affil[c]{Lowell Center for Space Science and Technology, University of Massachusetts Lowell, Lowell, MA, USA}
\affil[d]{Department of Astronomy and Steward Observatory, University of Arizona, Tucson, AZ, USA}

\authorinfo{Further author information: (Send correspondence to K.Z.D or E.S.D)\\K.Z.D: E-mail: derbyk@arizona.edu, E.S.D: E-mail: douglase@arizona.edu}

\begin{document} 
\maketitle

\begin{abstract}

Extreme wavefront correction is required for coronagraphs on future space telescopes to reach 10\textsuperscript{-8} or better starlight suppression for the direct imaging and characterization of exoplanets in reflected light. Thus, a suite of wavefront sensors working in tandem with active and adaptive optics are used to achieve stable, nanometer-level wavefront control over long observations. In order to verify wavefront control systems, comprehensive and accurate integrated models are needed. These should account for any sources of on-orbit error that may degrade performance past the limit imposed by photon noise. 

An integrated model of wavefront sensing and control for a space-based coronagraph was created using geometrical raytracing and physical optics propagation methods. Our model concept consists of an active telescope front end in addition to a charge-6 vector vortex coronagraph instrument. The telescope uses phase retrieval to guide primary mirror bending modes and secondary mirror position to control the wavefront error within tens of nanometers. The telescope model is dependent on raytracing to simulate these active optics corrections for compensating the wavefront errors caused by misalignments and thermal gradients in optical components. Entering the coronagraph, a self-coherent camera is used for focal plane wavefront sensing and digging the dark hole. We utilize physical optics propagation to model the coronagraphy’s sensitivity to mid and high-order wavefront errors caused by optical surface errors and pointing jitter. We use our integrated models to quantify expected starlight suppression versus wavefront sensor signal-to-noise ratio. 

\end{abstract}


\section{Introduction}
\label{sec:intro}  % \label{} allows reference to this section

\begin{figure} [h!]
   \centering\includegraphics[height=5cm]{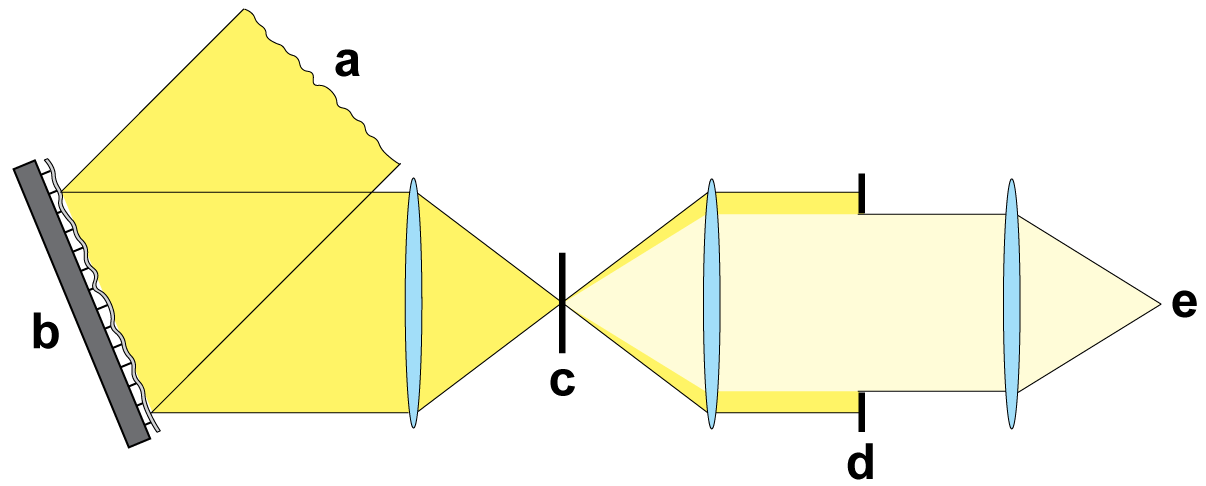}
   \caption{Basic coronagraph instrument layout. (a) An aberrated input wavefront is sent to (b) a deformable mirror (DM) to correct wavefront errors. The corrected wavefront is propagated to (c) a focal plane mask which diffracts on-axis starlight to the outer edges of the aperture where it is blocked by (d) a Lyot stop. Residual starlight and off-axis light from planets or debris disks makes it through the Lyot stop to (e) the final image plane.} 
   \label{fig:cgraph_layout} 
   \end{figure}

The existence of planets beyond our solar system was first confirmed in 1994\cite{wolszczan_confirmation_1994}. Since then, thousands more exoplanets have been discovered, primarily through radial velocity (RV) and transit measurements. These indirect methods detect exoplanets via the influence they exert over their host stars, making them ill-suited for detailed characterization. Direct imaging of exoplanets allows us to study them in far greater detail. By optically isolating planet light from starlight, exoplanets can be characterized at a much high signal-to-noise ratio (SNR) than possible with indirect methods. Additionally, direct imaging gives access to a larger sample of planets which are too small to exert significant influence over their host stars and planets that do not transit their host stars when viewed from Earth. However, direct imaging is incredibly difficult as exoplanets are orders of magnitude fainter than, and separated by only fractions of an arcsecond from their host stars. Thus, coronagraphs are used to suppress starlight and allow for exoplanet detection at small separations in reflected light. While coronagraphs come in many different flavors, most follow a similar scheme as shown in Figure \ref{fig:cgraph_layout}. These use a combination of pupil and focal plane masks to diffract on-axis starlight to the edges of the pupil, where it can be rejected using a Lyot stop while leaving off-axis planet light untouched\cite{zimmerman_lyot_2016, mawet_vector_2010, ruane_vortex_2018, laurent_apodized_2018, guyon_design_2018}.

Future observatories will be equipped with advanced coronagraphs with the goal of reaching 10\textsuperscript{-10} or higher contrast to survey large populations of exoplanets and search for signatures of life. Currently, 10\textsuperscript{-8} to 10\textsuperscript{-10} contrasts have been demonstrated in laboratory settings\cite{ashcraft_space_2022, van_gorkom_space_2022, trauger_coronagraph_2004, seo_testbed_2019, ndiaye_high-contrast_2013}, but significant work is needed to translate these to observatory operation. Even as contrasts dive deeper, limited photon flux remains a fundamental constraint when trying to detect small exoplanets. Large meter-class telescopes only collect a single photon every few seconds from such faint sources. Thus, long exposures over many hours of observation are required to build large amounts of data for post-processing in order to reach the threshold for detection. However, coronagraphs are incredibly sensitive to error, requiring picometer-level wavefront stability over these long exposures. Thus, adaptive optics (AO) are used for closed-loop wavefront sensing and control\cite{guyon_design_2018}. Coronagraph AO systems are typically comprised of wavefront sensors which feed a closed-loop control algorithm that commands a deformable mirror (DM). These algorithms iteratively optimize the DM shape to minimize the amount of residual starlight within a specified region of interest within the science image where we hope to find planets, a process known as "digging the dark hole"\cite{giveon_electric_2007, giveon_broadband_2007, haffert_implicit_2023, baudoz_self-coherent_2005, gerard_fast_2018, potier_comparing_2020, pueyo_optimal_2009, groff_methods_2015}.

Due to the extreme wavefront stability required to reach and maintain the deepest contrasts, a space-based observatory is needed as the presence of a turbulent, boiling atmosphere limits seeing for observatories on the ground. Still, space telescopes must contend with a plethora of on-orbit degradations. This complicates the modeling process significantly as these errors cover entirely different regimes which approximate the behavior of light. Almost all optical designs begin with geometric raytrace analysis. In this regime, the optical performance for a telescope and its instruments can be optimized to the diffraction limit. As its name suggests, geometric raytracing is ideal for modeling the low-order ray-based errors. This includes field-dependent aberrations and optical component misalignments which are a lingering threat on-orbit due to constantly changing thermal gradients which slowly warp and deform the spacecraft via thermal expansion and contraction\cite{blaurock_structural-thermal-optical_2005}. In addition, geometric raytracing is useful for modeling active optics which rely on the bulk motion of optical elements to correct low-order wavefront errors (WFE) caused by optical misalignments. This becomes a powerful tool for correcting errors such as defocus and coma which are typically some of the largest in terms of peak-to-valley (PV) WFE. Such high PV WFE would consume most, if not all of the dynamic range of a DM. 

Though raytrace analysis is a powerful tool in the ray-based regime, it can only get you to the diffraction limit. As mentioned previously, coronagraphs are incredibly sensitive instruments which respond to wavefront aberrations well below this limit. This ends up including effects such as optical surface errors, beamwalk on intermediate optics, telescope pointing jitter, or polarization aberrations\cite{mendillo_optical_2017, mendillo_polarization_2019}. Thus, physical optics propagation is a necessity when modeling their performance. Here, the wave-based effects of diffraction take over, requiring Fresnel and Fraunhofer integrals to model the physical optics propagation of light through an optical system. However, these integrals need to use the paraxial approximation to be accurate, which also breaks some of the assumptions used for geometric raytracing. This fundamental incompatibility between modeling light in these two regimes needs integrated optical models which are capable of bridging this gap.

In this manuscript, we present an integrated model of the 6.5-meter space telescope described in Douglas et al. of these proceedings. We utilize this model to simulate the performance of the observatory's potential wavefront sensing and control loops. First, a wide-field context camera (CC) placed at the image plane plane of the telescope which is used for phase retrieval to drive active primary and secondary mirrors. This telescope-corrected wavefront is then fed to our science instrument of interest: a charge-6 vector vortex coronagraph (VVC). This uses a self-coherent camera (SCC) to drive a 34 $\times$ 34 actuator DM for digging and maintaining a dark hole. We use our model to characterize the wavefront sensing performance and demonstrate closed-loop control of these instruments in simulation.

\section{Integrated Modeling of a Space-Based Observatory}
\label{sec:model}

An integrated model of a conceptual space-based observatory and its wavefront control loops was created using a combination of geometric raytrace analysis and physical optics propagation. Geometric raytrace analysis was performed using Zemax OpticStudio (ZOS) and its application programming interface (API). Physical optics propagation was performed using a combination of Physical Optics Propagation in Python (POPPY) and High Contrast Imaging in Python (HCIPy), two open-source Python libraries developed by teams of astronomers at the Space Telescope Science Institute and Leiden University, respectively\cite{perrin_poppy_2016, por_high_2018}.

Much of our error modeling efforts rely on the use of power spectral densities (PSD), which provide the spatial or temporal frequency spectra for a measurement. PSDs provide information for creating statistically representative synthetic errors and how they might evolve in time\cite{lumbres_extreme_wavefront_2022, douglas_laser_2019}. They allow for a convenient method of approximating realistic errors prior to obtaining real world measurements and are already commonly used for generating synthetic optical surface errors in physical optics propagation packages such as POPPY or HCIPy. As a first-approximation, the errors in our physical optics models are generated using simplified Von-Karman-like PSDs:

\begin{equation}
    PSD(f) = \frac{\beta}{(1+f/f_n)^\alpha}.
\end{equation}

where $f_n$ is the "knee frequency", $\alpha$ is the slope of the power law, and $\beta$ is a normalization factor.

\subsection{Telescope and Context Camera}
\label{sec:model_cc}

\begin{figure} [h!]
   \centering\includegraphics[height=6.5cm]{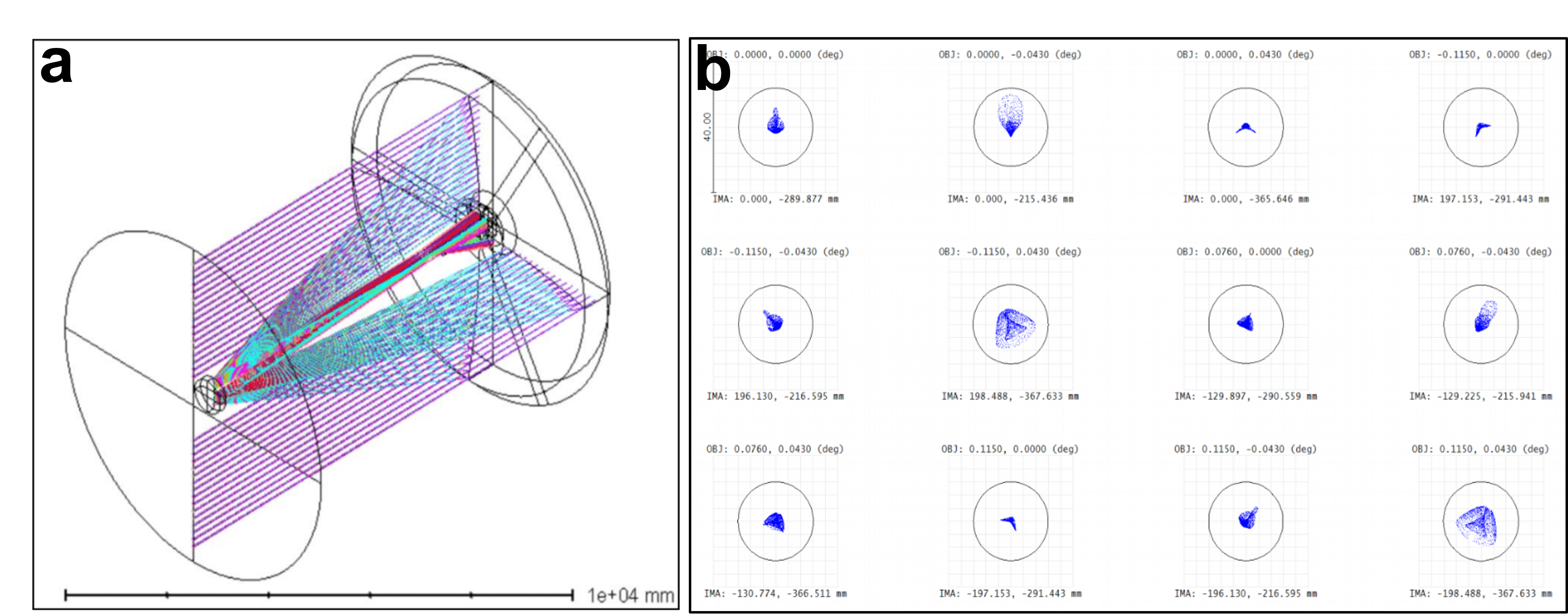}
   \caption{The on-axis TMA telescope design from Kim et al. of these proceedings\cite{kim_design_2023}. (a) The ZOS optical layout shows different fields being traced as different colors in addition to the shadow of our assumed secondary support structure on the primary mirror. (b) A full-field spot diagram shows diffraction limited performance across the FOV, the black circles show the Airy radius at 650 nm.} 
   \label{fig:telescope_design} 
   \end{figure}

We assume a wide-field context camera (CC) located at the focal plane of the telescope assembly. The telescope is assumed to be an on-axis compact TMA as discussed in Kim et al. of these proceedings\cite{kim_design_2023}. Figure \ref{fig:telescope_design} shows the ZOS optical layout in addition to a spot diagram demonstrating diffraction limited performance across our FOV. Being an on-axis design, we assume a secondary support structure comprised of three legs measuring 0.15-meters wide to support a 1.3-meter diameter secondary mirror to include in both our raytrace and physical optics models.

\begin{figure} [h!]
   \centering\includegraphics[height=8.75cm]{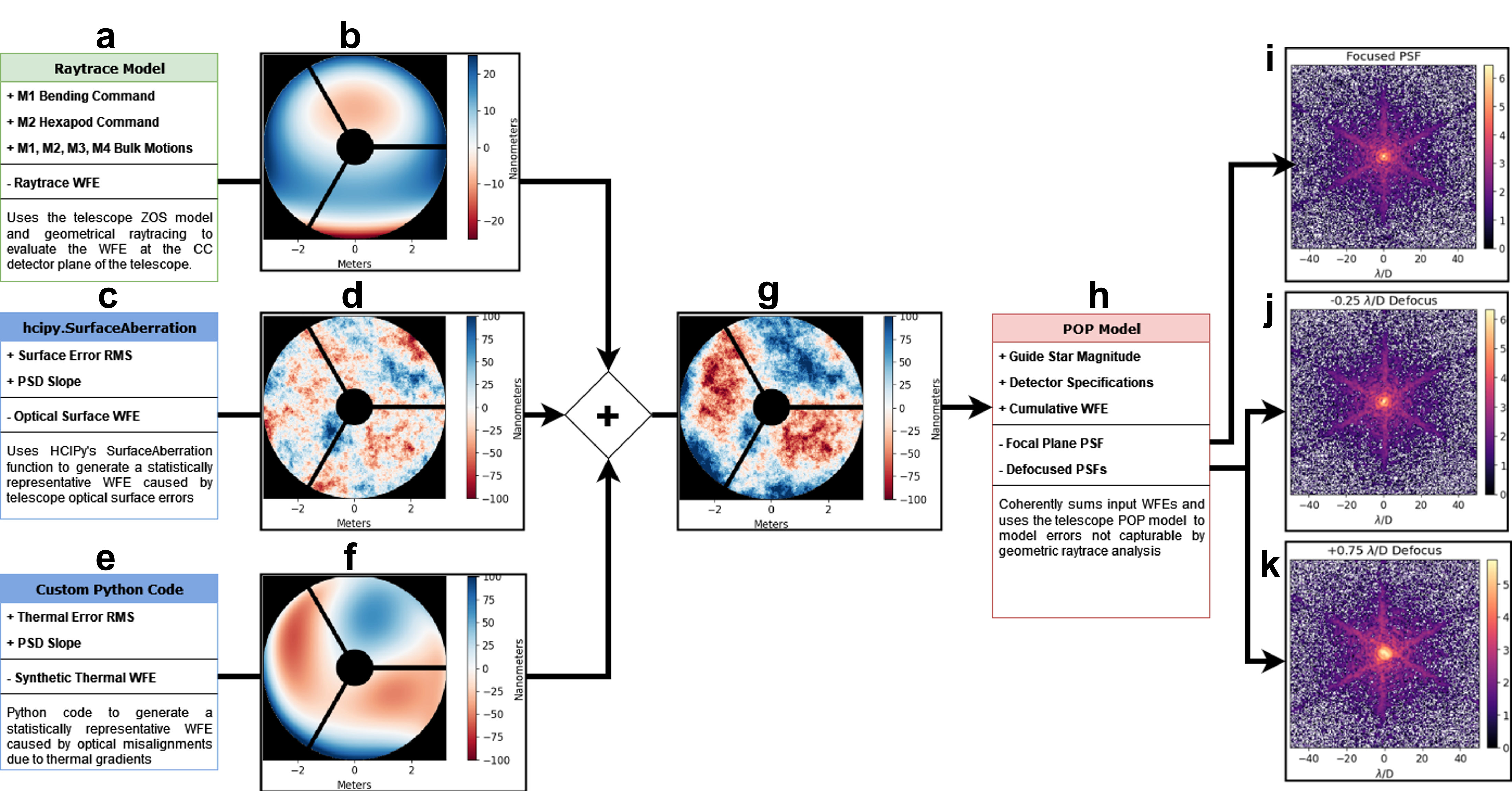}
   \caption{Telescope and CC simulation flow. (a) The raytrace model is used to evaluate (b) the raytrace WFE, (c) HCIPy's SurfaceAberration function is used to generate (d) the optical surface WFE, and (e) custom Python code is used to generate (f) the synthetic thermal WFE. These are coherently summed to obtain (g) the cumulative WFE and fed to (h) the physical optics propagation (POP) model which simulates (i) the focal plane PSF and (i, k) two defocused PSFs for the CC.} 
   \label{fig:telescope_sim_flow} 
   \end{figure}

The simulation flow for the CC is rather involved due to the use of both geometrical raytracing and physical optics propagation as shown in Figure \ref{fig:telescope_sim_flow}. It begins in the ZOS raytrace model where optical element bulk motions can be provided via a custom ZOS-API script. These bulk motions could represent misalignments caused by thermal gradients or in the case of the secondary mirror (M2), the correction being fed to the M2 position. The ZOS-API is then used to execute a batch raytrace of the telescope at a user specified field location and the image plane WFE is extracted (hereafter referred to as the raytrace WFE). This comprises one of three wavefront maps which will be coherently summed in the physical optics model. The second is a 43 nm root-mean-square (RMS) phase mask generated using HCIPy's SurfaceAberration function to represent optical surface errors (hereafter referred to as optical surface WFE). This function generates a statistically representative surface error from a PSD with a slope of -2.5. 43 nm RMS was assumed to match the optical surface errors allotted to the telescope optics in the error budget discussed Choi et al. of these proceedings\cite{choi_approaches_2023}. The final wavefront map we referred to as synthetic thermal WFE. This was generated from a Von-Karman-like PSD with a slope of 6 and a knee-frequency placed at \~Zernike 5 (Noll-indexed) using custom Python code. This sharp roll-off at a rather low spatial frequency was assumed because thermal errors primarily manifest as optical misalignments which lead to low-order errors. Higher-orders were left in as we expect thermal errors to cause additional optical surface deformations on each mirror.

The raytrace, optical surface, and synetic thermal WFE were coherently summed into what we refer to as the cumulative WFE. Note that this cumulative WFE is another important output of the CC model which feeds into the (VVC) model which is discussed further in Section \ref{sec:model_vvc}. This was fed to the physical optics model of the telescope which propagated the cumulate WFE to the CC detector plane using Fraunhofer propagation to give a focal plane point spread function (PSF). In addition, the focal plane PSF was propagated to two additional defocus planes near the CC image plane using angular spectrum propagation. This gave two additional defocused PSFs which are used for focus diverse phase retrieval (FDPR) in the CC wavefront sensing and control loop discussed in Section \ref{sec:model_wfsc}. At the detector plane, we apply 10 milli-arcsecond RMS jitter by convolving the image with a Gaussian kernel. For the detector, we assumed 0.002 electrons/pixel/s dark current and 1.4 electrons/read of read noise. This corresponds to a complementary metal oxide semiconductor (CMOS) detector operating at a high gain in a thanks to commercial low-noise CMOS detectors being capable tiling large FOVs at a relatively low cost.

\subsection{Vector Vortex Coronagraph}
\label{sec:model_vvc}

We assume that the on-axis wavefront from the telescope is able to pass through the CC detector plane to enter our vector vortex coronagraph (VVC) instrument. We passed through the on-axis wavefront because that is where the field dependent aberrations  are minimal. Upon entering the VVC, we assume that he full aperture of the telescope is not used. Instead, a pupil mask is used to select smaller sub-apertures between the secondary support struts. This is because VVCs are notoriously difficult to work with when using segmented or obscured apertures\cite{mawet_vector_2010-1, ruane_performance_2017}. We assume a 2.4-meter sub-aperture for our model, which is roughly 95\% of the largest sub-aperture available for our telescope with its assumed M2 and secondary support structure.

   \begin{figure} [h!]
   \centering\includegraphics[height=9cm]{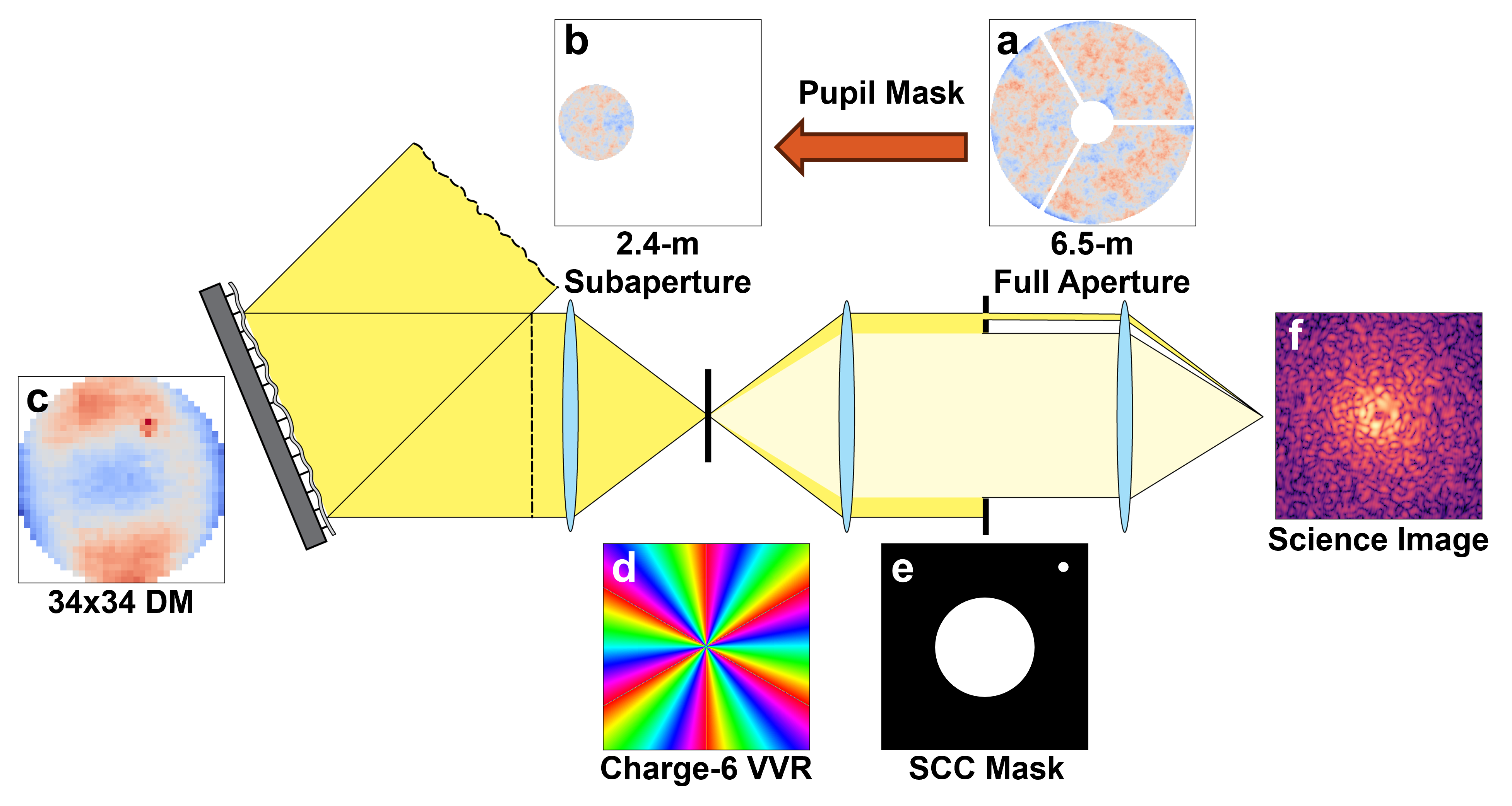}
   \caption{A diagram showing the the layout of our VVC model. (a) The 6.5-m residual telescope wavefront is assumed to be masked down to (b) an unobscured 2.4-m sub-aperture. The aberrated wavefront is first sent to (c) a 34 $\times$ 34 actuator DM for wavefront correction. The corrected wavefront is then focused onto (d) a charge-6 VVR which sends on-axis starlight to the edges of the geometric pupil where it is rejected by (e) the SCC Lyot Stop. Residual starlight and off-axis planet light makes it through the SCC Lyot stop to (f) the detector plane.} 
   \label{fig:vvc_model} 
   \end{figure} 

For the VVC, we used the physical optics model of the University of Arizona Space and Astrophysics Lab's (UASAL) Space Coronagraph Optical Benchtop (SCoOB) model built using POPPY\cite{ashcraft_space_2022, van_gorkom_space_2022}. SCoOB is a vacuum-compatible charge-6 VVC testbed which uses a 32 $\times$ 32 actuator Boston Micromachines Kilo-DM for wavefront correction. We use the SCoOB model as a stepping stone for adapting our model to a flight-like system being proposed for NASA's Habitable Worlds Observatory, allowing us to test and validate our wavefront sensing and control loops in the lab. The only modification to the model is the aforementioned pupil mask which selects a circular, unobscured sub-aperture between the secondary support struts present in the cumulative WFE map discussed in \ref{sec:model_cc}. This sub-aperture wavefront is then relayed to the DM used for wavefront correction and dark hole digging. Following the DM, the wavefront is slowly focused at $\sim$f/60 onto a charge-6 vector vortex retarder (VVR) at the coronagraph focal mask plane and then re-collimated. The VVR sends on-axis starlight to the edges of the sub-aperture where it is then rejected by a self-coherent camera (SCC) Lyot stop in the downstream pupil plane. The SCC Lyot stop was set to 90\% the diameter of the sub-aperture with a pinhole measuring 3\% the diameter of the sub-aperture. Finally, after the starlight is rejected, the wavefront is focused on the science camera located at the final VVC image plane. At the image plane, we apply 10 milli-arcsecond RMS jitter by convolving the image with a Gaussian kernel. For the detector, we again assumed 0.002 electrons/pixel/s dark current and 1.4 electrons/frame of read noise.

\subsection{Closed-Loop Wavefront Sensing and Control}
\label{sec:model_wfsc}

The CC contains the telescope's wavefront sensing and control loop. Here, bright stars in the FOV are used to perform focus-diverse phase retrieval (FDPR). FDPR is a method of recovering wavefront phase using PSF intensity images with known amounts of injected defocus diversity\cite{fienup_phase_1982, dean_phase_2006}. It has been successfully demonstrated on-orbit in diagnosing the Hubble Space Telescope's mirror defect and during the commissioning of the James Webb Space Telescope. We use an algorithmic differentiation method as described in Jurling and Fienup\cite{jurling_applications_2014}.

   \begin{figure} [h!]
   \centering\includegraphics[height=15cm]{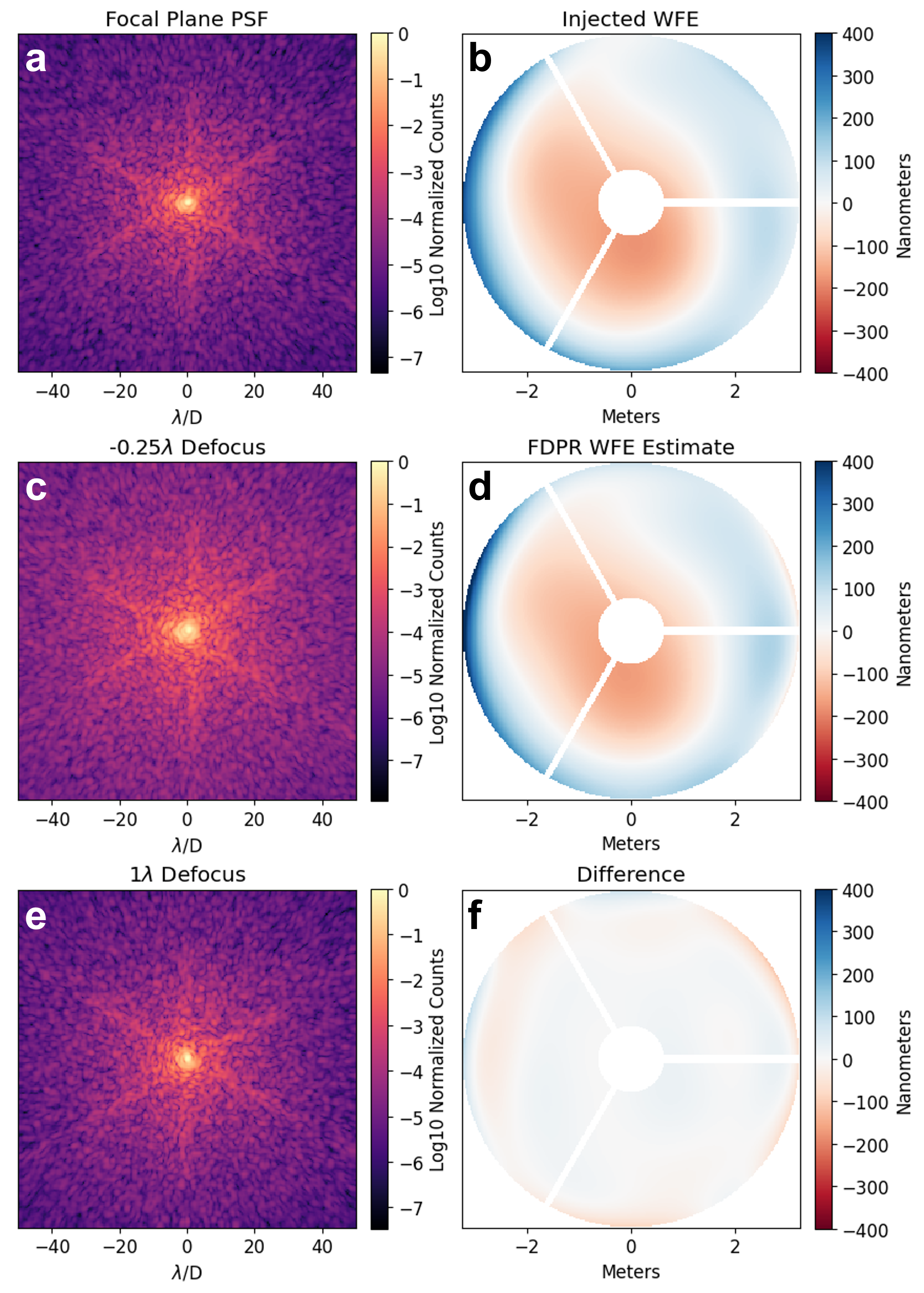}
   \caption{FDPR demonstration for our telescope and CC instrument concept. The (a) focal plane PSF is shown for (b) 100 nm RMS injected WFE. The (c, e) defocused PSFs are fed to our FDPR algorithm to return (d) the FDPR WFE estimate. We also show (f) the difference between injected and estimated WFE.} 
   \label{fig:fdpr_example} 
   \end{figure} 

FDPR estimations of wavefront error at the focal plane are used to control an active primary (M1) and secondary mirror (M2). M1 is assumed to use a set of actuators to deform the surface shape into a set of bending modes which can correct low- to mid-order spatial errors up to roughly Zernike 37 (using Noll indexing). Blomquist et al. in these proceedings gives a detailed explanation of the M1 correction mechanism and its bending modes\cite{blomquist_analysis_2023}. Meanwhile, M2 is assumed to have 5 DOFs corresponding to x, y, and z position in addition to x and y tilt. M1 and M2 correction is calculated and applied using custom MATLAB code before being fed back into the rest of the CC simulation loop.

   % \begin{figure} [h!]
   % \centering\includegraphics[height=3.5cm]{figures/0?_wfsc_flow.png}
   % \caption{MAKE FLOW CHART OF WFSC LOOPS BEING RUN IN THE CC AND VVC} 
   % \label{fig:wfsc_flow} 
   % \end{figure} 

The VVC contains its own wavefront sensing and control loop for suppressing mid- to high-order spatial frequencies which manifest as speckles at the VVC detector plane. We assumed a self-coherent camera (SCC) as a focal plane wavefront sensor, the algorithm for which can again be found in Lina. The SCC is a wavefront sensing method capable of sensing mid- to high-spatial frequency wavefront errors to either dig or continuously maintaining a dark hole\cite{potier_comparing_2020, mazoyer_high-contrast_2014, galicher_self-coherent_2010, galicher_wavefront_2008, baudoz_self-coherent_2005}. It operates by introducing an offset pinhole at the Lyot stop, allowing a small amount of stellar light through to the science image. This stellar "reference channel" will interfere with coherent stellar speckles, while leaving incoherent planet light untouched. This difference in coherence also opens possibilities for post-data acquisition image processing using coherent differential imaging (CDI)\cite{bottom_speckle_2017, gerard_fast_2018, jovanovic_review_nodate}.

   \begin{figure} [ht]
   \centering\includegraphics[height=15cm]{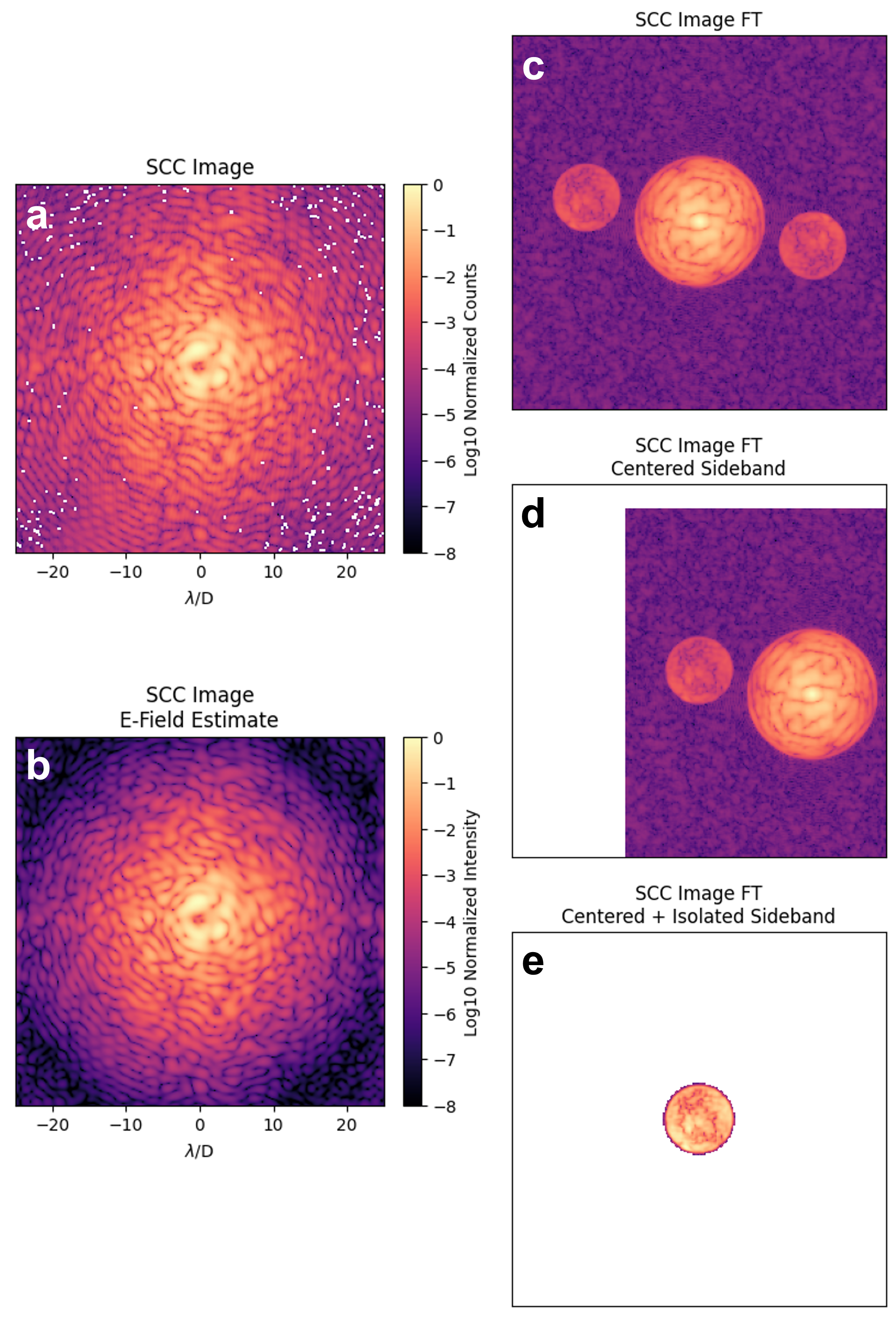}
   \caption{SCC demonstration estimating the electric field at the detector plane. (a) The SCC science image shows the actual intensity of the electric field at the detector plane. A FT of the image reveals (c) 3 bands in frequency space. (d) A sideband can be shifted to the center and (e) isolated. The inverse FT of this centered and isolated sideband gives an estimate of the electric field at the detector plane. We show (b) the intensity of this electric field estimation.} 
   \label{fig:scc_example} 
   \end{figure} 

The Fourier Transform (FT) of an SCC image reveals a large central band with two smaller side-bands. These will not overlap if the separation between the pinhole and Lyot stop is greater than 1.5 times the diameter of the Lyot stop and pinhole added together. In this case, we can center and isolate one of the side-bands. The inverse FT of this isolated side-band returns an estimate of the electric field at the detector plane using only a single science image. This estimation can then be used for wavefront correction by controlling a DM at an upstream pupil plane from the VVR. The ability to estimate the field using a single image just by introducing a pinhole at the Lyot stop makes the SCC an attractive and elegant solution for faster dark-hole digging and maintenance on future space-based coronagraphs.

\section{Simulation Results}
\label{sec:sims}

We use our integrated models of the telescope, CC, and VVC described in Section \ref{sec:model} to demonstrate closed-loop control in simulation. Section \ref{sec:sims_cc} discusses FDPR performance as a function of SNR in addition to showing closed-loop control of the telescope and CC. Section \ref{sec:sims_vvc} demonstrates dark hole digging using the VVC.

\subsection{Context Camera Performance}
\label{sec:sims_cc}

Ultimately, our control loops are fundamentally limited by the wavefront sensing architecture in addition to detector and photon noise. Thus, our first goal is to find and characterize this limit. We show FDPR estimate error for the CC as a function of RMS wavefront error and the focal plane PSF SNR. Focal plane PSF SNR allows our results to stay "model agnostic" so that others working on different systems have a method of comparison with this work. We define PSF SNR as the ratio between the peak PSF intensity versus the average background noise of a dark image. A single synthetic thermal WFE was generated from the Von-Karman-like PSD described in Section \ref{sec:model_cc} and was scaled to obtain multiple WFE RMS values. Note that this injected error also error rides on top of the 43 nm RMS optical surface WFE present in the CC model. For this simulation, two PSFs defocused by -0.25 and 1 $\lambda$ were used for FDPR. Our FDPR algorithm estimates the first 37 Zernikes (Noll-indexed) of wavefront error at the CC plane. We define RMS estimate error as the RMS difference in the actual wavefront error versus the FDPR estimated wavefront error at the CC plane.

   \begin{figure} [ht]
   \centering\includegraphics[height=9cm]{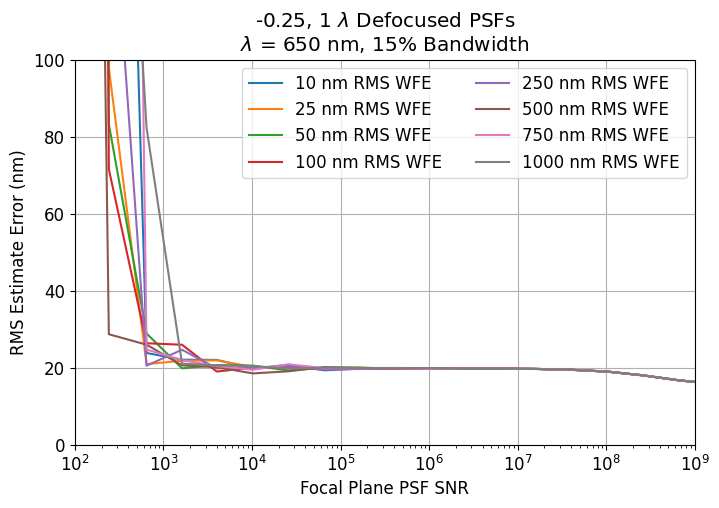}
   \caption{FDPR RMS estimate error versus focal plane PSF SNR for several injected RMS WFE values.} 
   \label{fig:fdpr_vs_snr} 
   \end{figure} 

We find that our FDPR algorithm quickly converges to an estimate error floor of 20 nm RMS above a focal plane PSF SNR of 500. For our 6.5-meter aperture and assuming a 60 second exposure time, this corresponds to roughly a 20th magnitude star - of which we expect several in the CC FOV during observations. With further optimizations to the algorithm, we expect to lower this estimate error floor below 10 nm RMS. Our FDPR algorithm has a large dynamic range and achieves similar performance for RMS WFE values across 2 orders of magnitude. Since the thermal errors we expect to see on-orbit evolve slowly, we expect this to be more than sufficient for correction even if our closed control loop updates at slower intervals on the order of several minutes. Currently, our simulations do not account for field-dependent aberrations since the guide star PSF is assumed to be on-axis. Future work will use multiple guide star PSFs across the FOV to estimate and correct the on-axis field, as that is where instruments are likely to be placed to mitigate field-dependent aberrations.

\begin{figure} [ht]
   \centering\includegraphics[height=8cm]{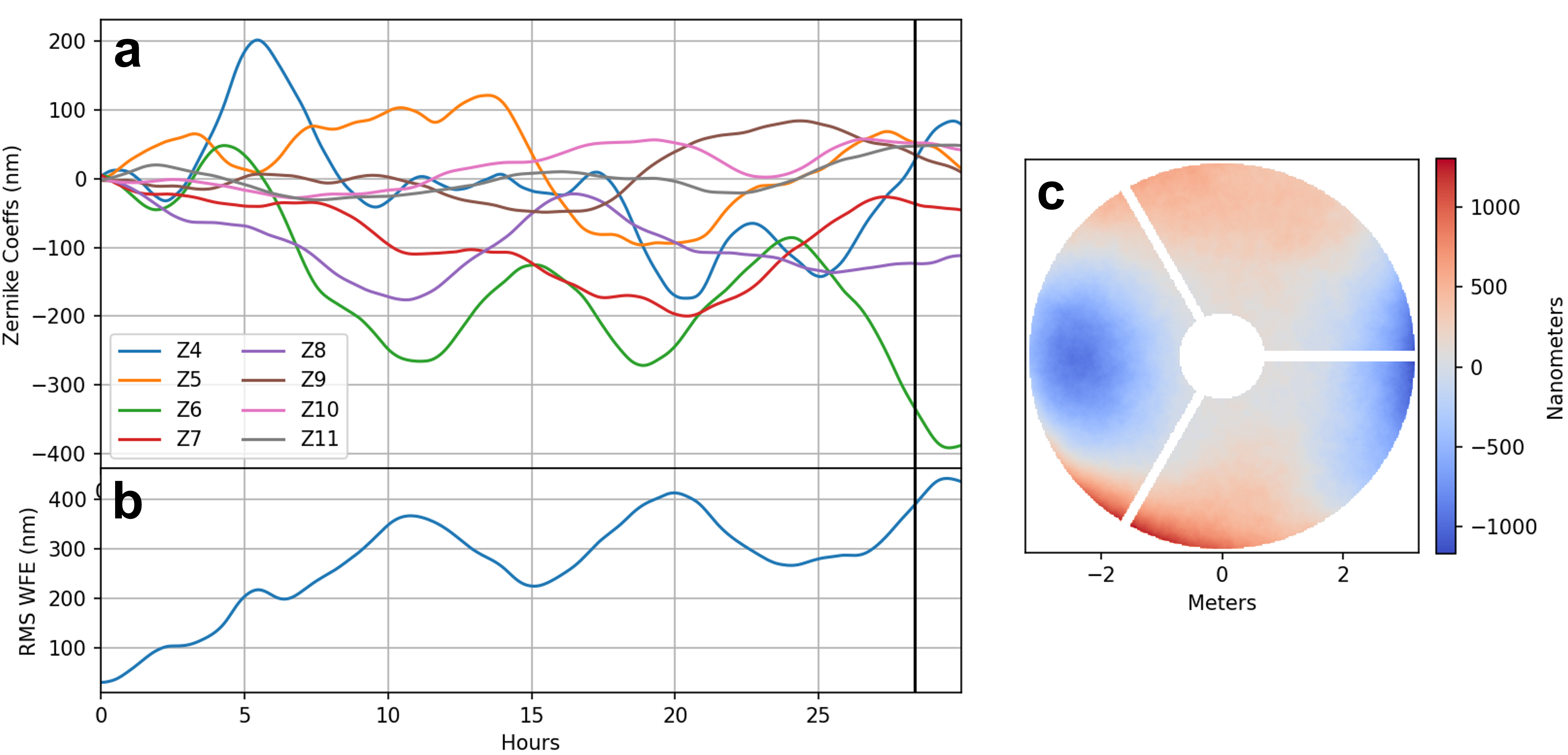}
   \caption{Telescope and CC performance without closed loop control. (a) Zernikes 3 through 11 of the synthetic thermal WFE evolve slowly over time and degrade (b) the cumulative RMS WFE significantly as a result. (c) A representative cumulative WFE error map corresponding to the vertical black line in the plots is also shown.} 
   \label{fig:fdpr_closed_loop_pre} 
   \end{figure}

Assuming a 17th magnitude on-axis guide star, we demonstrate closed-loop control of the telescope and CC across a 30 hour time series in simulation. For this simulation, we assume a static optical surface WFE and a dynamic synthetic thermal error. The time evolution of the synthetic thermal WFE was assumed to follow another Von-Karman-like PSD with a slope of 7 and knee frequency placed at 0.45-hour\textsuperscript{-1}. This temporal PSD is normalized so that Zernike 4 changes at a rate of 1 nm/minute and falls off quickly as we expect thermal gradients to evolve slowly in time. Douglas et al. of these proceedings shows the temporal PSDs of the first 25 Zernike coefficients used for generating the synthetic thermal error in these simulations\cite{douglas_big_risks_2023}. We assumed each term started from zero and was independently generated. This ends up being a conservative assumption as a real world system would have correlated Zernike terms which an optimized controller could take advantage of. Finally, 10 mas pointing jitter was added by convolving with a Gaussian Kernel. Figure \ref{fig:fdpr_closed_loop_pre} shows how each Zernike term (37 are shown because M1 and M2 can correct up to \~Zernike 37) of the synthetic thermal WFE evolves over the 40 hour time period. In addition, a representative cumulative WFE map and a plot of cumulative WFE RMS over time is shown. 

We assume 60 second detector exposures allowing our control loop operate at 0.0167 Hz. This consists of defocused PSF images being first sent to our FDPR algorithm to estimate the on-axis WFE. The on-axis WFE is then sent to a controller which sets the M2 position to correct for defocus and coma while other aberrations are corrected by M1's actuated bending modes.

   \begin{figure} [ht]
   \centering\includegraphics[height=8cm]{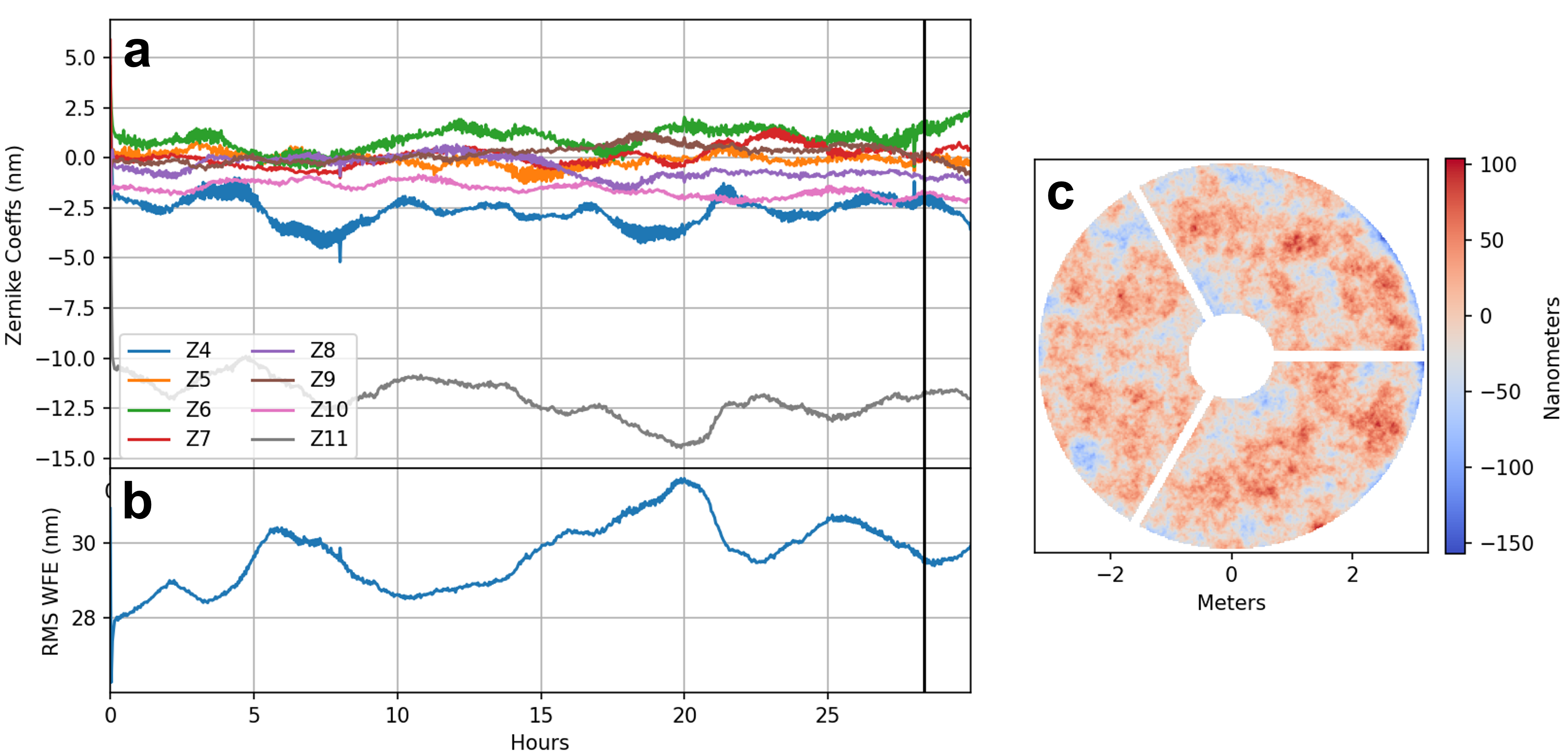}
   \caption{Telescope and CC performance with closed loop control. (a) Zernikes 3 through 11 of the synthetic thermal WFE and (b) the cumulative RMS WFE are significantly more stable over time. (c) A representative cumulative WFE error map corresponding to the vertical black line in the plots is also shown.} 
   \label{fig:fdpr_closed_loop_post} 
   \end{figure} 

Figure \ref{fig:fdpr_closed_loop_post} shows the evolution of each Zernike term of the synthetic thermal WFE after closed-loop correction. Again, a representative cumulative WFE map and a plot of cumulative WFE RMS over time is shown. With closed-loop control, the cumulative WFE RMS hovers around 30 nm for nearly the entire time period which will allow for sub-nanometer wavefront control in our VVC to reach 10\textsuperscript{-8} contrasts. We also show that the cumulative WFE RMS is stabilized significantly, which enables longer duration exposures on faint objects without large changes in the PSF. We can see that Zernike 11 (e.g. 1st order spherical) is not well corrected and constitutes most of the residual error. This is because spherical aberration is characterized by a sharp change in shape near the edges of the pupil, which is difficult to correct using an actuated primary mirror without putting significant stress on the optic. Future work will perform more simulations using multiple PSD realizations to establish a bounding case for specifying closed-loop control as a function of the PSD. In addition, we will add the correlations between Zernike terms expected of a real-world observatory and optimize our controller accordingly.

\subsection{Vector Vortex Coronagraph Performance}
\label{sec:sims_vvc}

\begin{figure} [ht]
   \centering\includegraphics[height=7cm]{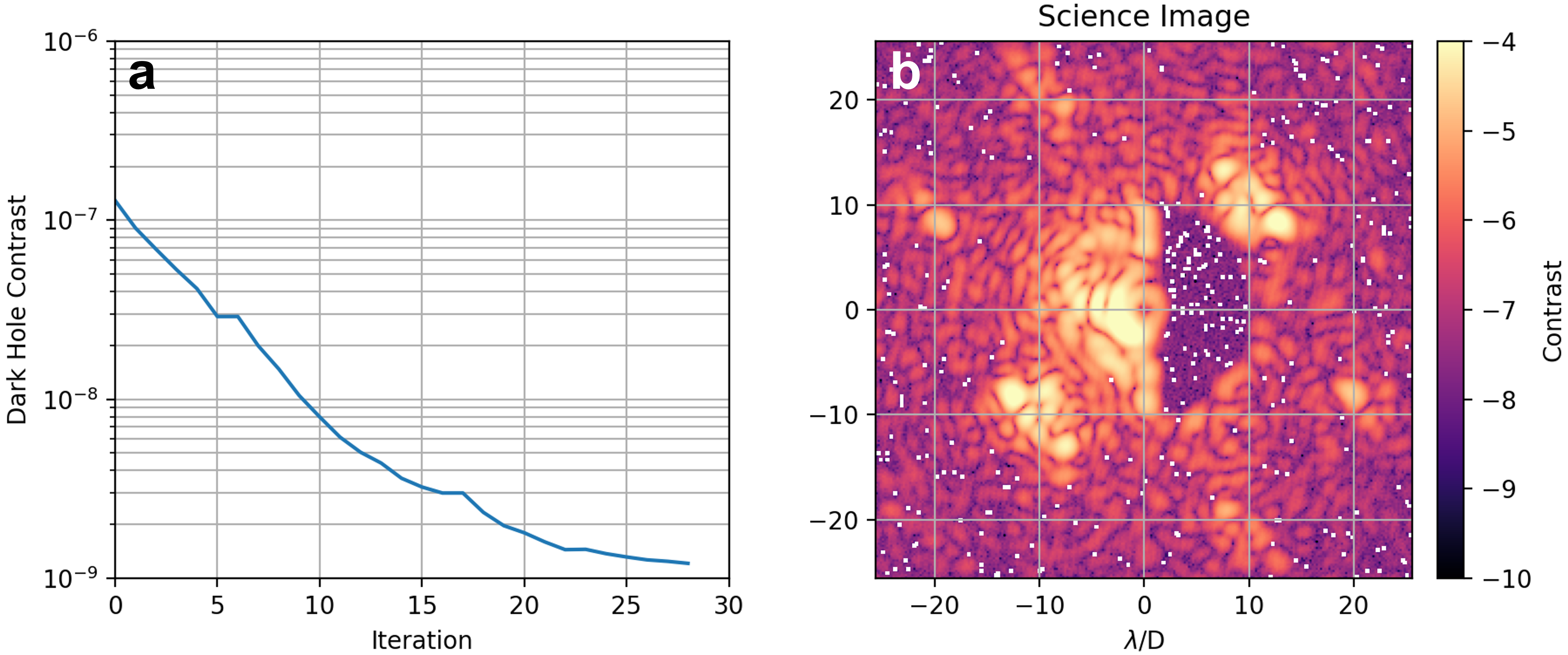}
   \caption{Dark hole digging using our VVC model employing an SCC for focal plane wavefront sensing. (a) The dark hole contrast over 40 iterations shows fast convergence down to a final contrast of 10\textsuperscript{-9}. (b) The final science image of the dark hole is shown as well.} 
   \label{fig:vvc_dark_hole} 
   \end{figure}

Assuming a magnitude 3 target star and 60 second exposures, we use our VVC model to demonstrate 10\textsuperscript{-9} contrast using an SCC as shown in Figure \ref{fig:vvc_dark_hole}. We assume 35 nm RMS of initial WFE caused by a combination of optical surface errors within the VVC and residual WFE from the telescope. Note that the residual 30 nm RMS WFE shown in the telescope closed-loop control simulation in Section \ref{sec:sims_cc} is calculated across the entire 6.5-m aperture. The VVC uses a 2.4-m sub-aperture, so the residual telescope WFE the VVC sees is reduced to roughly 20 nm RMS. The dark hole is dug in 40 iterations. We use beta regularization for generating the control matrix. 5 nm pokes were used for calibrating the Jacobian using Fourier modes.

   % \begin{figure} [ht]
   % \centering\includegraphics[height=3.5cm]{figures/0?_scc_vs_snr.png}
   % \caption{CONTRAST VS SNR} 
   % \label{fig:scc_vs_snr} 
   % \end{figure} 

Again, our control loops are fundamentally limited by the wavefront sensing architecture in addition to detector and photon noise, so we found this limit using our VVC model. We show the dark hole contrast as a function of dark hole speckle SNR. Dark hole speckle SNR again allows our results to stay model agnostic. We define dark hole speckle SNR as the ratio of average speckle intensity in the dark hole versus the average background noise of a dark image. We use the exact same method of dark hole digging as explained previously for demonstrating 10\textsuperscript{-9} contrast. We define dark hole contrast as the average dark hole contrast after 40 iterations of digging with the SCC.

   \begin{figure} [ht]
   \centering\includegraphics[height=9cm]{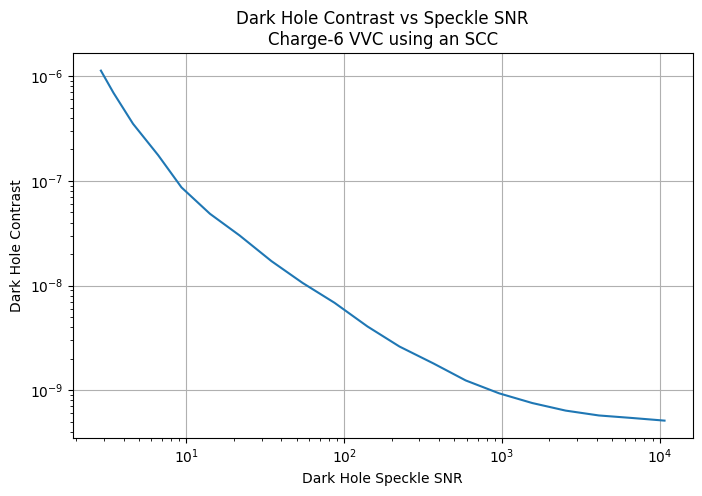}
   \caption{Final dark hole contrast values versus dark hole speckle SNR.} 
   \label{fig:vvc_vs_snr} 
   \end{figure} 

We find that our VVC outfitted with an SCC is capable of reaching sub-10\textsuperscript{-8} contrasts for a speckle SNR above 50. For our 2.4-meter sub-aperture and assuming a 60 second exposure time, this corresponds to roughly a 6th magnitude star. Crucially, this is sufficient for several nearby interesting target stars such as Alpha Cen or Eps Eri. However, in the case of Alpha Cen, binary star wavefront control will also need to be demonstrated in future work as the speckles from its companion will significantly degrade contrast. Currently, our VVC simulations do not account for beamwalk on intermediate optics or polarization aberrations. While we do not expect these to significantly degrade performance, future work will need to verify this and characterize these degradations. 

\section{Conclusion}
\label{sec:conclusion}

We have created integrated models of an example 6.5-meter space telescope equipped with CC and VVC instruments using a combination of geometric raytracing and physical optics propagation. These models are capable of approximating common on-orbit degradations such as optical surface errors, pointing jitter, and changing thermal gradients using PSDs. We use these approximations to characterize the expected performance of our instruments and demonstrate closed-loop control in simulation. We find that our telescope concept and CC instrument concept is capable of providing FDPR estimates of the telescope WFE for 20th magnitude and brighter guide stars. We also demonstrate closed loop control of the telescope and CC instrument using an active M1 and M2 for a 17th magnitude guide star. In addition,   we find our VVC instrument concept is capable of reaching 10\textsuperscript{-9} and 10\textsuperscript{-8} dark hole contrast for 3rd and 6th magnitude target stars, respectively using an SCC for focal plane wavefront sensing.

Future work will add support for modeling beamwalk on intermediate optics and polarization raytracing to both models, as we expect these degradations to play a role in setting the contrast limit of such systems. In addition, we hope to significantly increase the computational speed and efficiency of our models by migrating portions currently in ZOS-API and MATLAB to Python, where we can take advantage of GPU acceleration built and pre-existing open source packages. With these improved models, we will establish bounding cases for closed-loop control in both instruments as a function of the various degradation sources. This will allow us to define requirement envelopes parametrically.

\acknowledgments % equivalent to \section*{ACKNOWLEDGMENTS}     

Portions of this research were supported by funding from the Technology Research Initiative Fund (TRIF) of the Arizona Board of Regents and by generous anonymous philanthropic donations to the Steward Observatory of the College of Science at the University of Arizona.
 
% References
\bibliography{zotero_full} % bibliography data in report.bib
\bibliographystyle{spiebib} % makes bibtex use spiebib.bst

\end{document}